\UseRawInputEncoding

\documentclass{ieeeaccess}

\usepackage[T1]{fontenc}
\usepackage[utf8]{inputenc}

\usepackage{amsmath,amssymb,amsfonts,bm}

\usepackage{graphicx}
\usepackage{subfigure}

\usepackage{algorithm}
\usepackage{algorithmic}

\usepackage{multirow,tabularx,booktabs}

\usepackage{textcomp,microtype,anyfontsize,lettrine}
\usepackage{color,soul}

\usepackage{epsf,epsfig,epic,eepic}

\usepackage{cite}

\usepackage{setspace}

\def \max{\operatornamewithlimits{max}}

\def \arg{\operatorname{arg}}

\def \beqi{\begin{IEEEeqnarray}{rcl}\IEEEyesnumber}
\def \eeqi{\end{IEEEeqnarray}}

\def \beq{\begin{equation}}                     \def \eeq{\end{equation}}
\def \beqn{\begin{eqnarray}}                    \def \eeqn{\end{eqnarray}}
\def \bmat{\begin{bmatrix}}                     \def \emat{\end{bmatrix}}
\def \bmats{\left[\begin{smallmatrix}}          \def \emats{\end{smallmatrix}\right]}

\renewcommand{\eqref}[1]{(\ref{#1})}

\allowdisplaybreaks
\def\BibTeX{{\rm B\kern-.05em{\sc i\kern-.025em b}\kern-.08em
    T\kern-.1667em\lower.7ex\hbox{E}\kern-.125emX}}

\begin{document}

\history{Date of publication xxxx 00, 0000, date of current version xxxx 00, 0000.}
\doi{10.1109/ACCESS.2017.DOI}

\title{Design of Orthogonal Phase of Arrival Positioning Scheme Based on 5G PRS and Optimization of TOA Performance}

\author{\uppercase{Hyejin Shin}\authorrefmark{1} \IEEEmembership{Student Member, IEEE},
\uppercase{Sohee Kim}\authorrefmark{1} \IEEEmembership{Student Member, IEEE},
\uppercase{Ilmu Byun}\authorrefmark{2,*} \IEEEmembership{Member, IEEE},
and
\uppercase{Juyeop Kim}\authorrefmark{1,*} \IEEEmembership{Member, IEEE}
}

\address[1]{Department of Electrical Engineering, Sookmyung Women\'s University, Seoul, South Korea}
\address[2]{Korea Railroad Research Institute, Uiwang, South Korea}

\corresp{*Corresponding authors: Juyeop Kim (e-mail: jykim@sookmyung.ac.kr) and Ilmu Byun (e-mail: ilmubyun@krri.re.kr).}

\tfootnote{This research was supported by a grant from R\&D Program (Private 5G-Railway core technology development for railway digital transformation, PK2404D1) of the Korea Railroad Research Institute and the Institute of Information \& Communications Technology Planning \& Evaluation (IITP) grant funded by the Korea government (MSIT) (No. 2021-0-00165, Development of 5G+ Intelligent Base Station Software Modem).}

\markboth
{H.~Shin \headeretal: Design of Orthogonal Phase of Arrival Positioning Scheme Based on 5G PRS and Optimization of TOA Performance}
{H.~Shin \headeretal: Design of Orthogonal Phase of Arrival Positioning Scheme Based on 5G PRS and Optimization of TOA Performance}

\begin{abstract}
This study analyzes the performance of positioning techniques based on configuration changes of 5G New Radio (NR) signals. In 5G networks, a terminal's position is determined from the Time of Arrival (TOA) of Positioning Reference Signals (PRS) transmitted by base stations. We propose an algorithm that improves TOA accuracy under low sampling-rate constraints and implement 5G PRS for positioning in a software-defined modem. We also examine how flexible time–frequency resource allocation of PRS affects TOA estimation accuracy and discuss optimal PRS configurations for a given signal environment.
\end{abstract}


\titlepgskip=-15pt

\maketitle

\section{Introduction}
Recently, research interest in mobile positioning has grown significantly due to its importance in applications such as remote control, target tracking, and vehicular navigation \cite{ref1y, ref2y}. 
Although the Global Navigation Satellite System (GNSS) has traditionally ensured high positioning accuracy in outdoor environments \cite{ref3y, ref4y}, it falls short of fulfilling the requirements of emerging services like IoT and remote operations in terms of accuracy, latency, and availability \cite{ref5y, ref6y, ref7y}.
It also faces challenges in providing accurate positioning in indoor environments \cite{ref8y,ref9y}.To overcome the limitations of GNSS, recent research has increasingly focused on leveraging cellular networks \cite{ref10y, ref11y, ref12y}.Wireless signals transmitted by nearby base stations can potentially lead to more accurate positioning compared to GNSS signals in many scenarios \cite{ref13y}.Several cellular-based positioning methods can be classified based on the types of information used to determine the location of mobile devices.Methods based on received signal strength (RSS), which estimate the relative distance based on the strength of received signals \cite{ref14y,ref15y}, cause significant errors due to the uncertainty of path loss \cite{ref16y}.
Angle-of-arrival (AOA)-based methods, which focus on the intersection of the lines of arrival directions \cite{ref17y,ref18y,ref19y}, require sophisticated antenna hardware. Unlike the other methods, the Time of Arrival (TOA) and Time Difference of Arrival (TDOA) based methods, which measure the time of signal arrival \cite{ref20y,ref21y, ref22y}, achieve high precision positioning based on accurate synchronization \cite{ref23y, ref24y}. 
Meanwhile, TOA and TDOA-based methods offer several advantages, but also face challenges such as multipath interference and signal blockage \cite{ref25y}. To address these issues, conventional studies have considered the estimation of the channel impulse response (CIR) or the channel frequency response (CFR).
These studies also suggest the allocation of multiple reference signals for accurate channel estimation in LTE and 5G \cite{ref26y, ref27y}. In addition, machine learning is jointly considered with fingerprinting \cite{ref29y}. Based on this approach, further studies integrate with beamforming and AOA estimation to develop hybrid solutions \cite{ref28y,ref30y}.Along with the growing interest in cellular-based positioning, long-term evolution (LTE) and 5G new radio (NR) embrace Time Of Arrival (TOA) and Time Difference Of Arrival (TDOA)-based methods from a standard perspective \cite{ref31y}. 
These systems provide positioning services through the positioning reference signal (PRS), whose configuration is flexible to various signal environments. 
LTE and 5G base stations are allowed to transmit PRS in various patterns, allowing a wide use of subcarriers within the frequency band and effectively avoiding interference between neighboring base stations \cite{ref32y,ref33y}. 
Hence, operators can easily provide positioning services by enabling PRS transmission and TOA and TDOA measurements by mobile devices.
Several studies have addressed positioning issues in terms of PRS in LTE and 5G NR \cite{ref34y}.
The effect of resource allocation patterns and subcarrier spacing configuration on positioning accuracy is analyzed at a simulation level \cite{ref35y,ref36y, ref37y}.
The deployment of base stations is considered for efficient positioning based on TDOA \cite{ref38y,ref39y}.
Positioning in 5G NR is further investigated in indoor or densely populated urban environments \cite{ref40y, ref41y}. 
Further studies in terms of indoor positioning provide improved algorithms within a few centimeters by mitigating multipath effects \cite{ref42y} and considering timing delay \cite{ref43y,ref44y}.In addition, fingerprint learning methods using neural network techniques are proposed to improve positioning accuracy \cite{ref45y}.
Traditional PRS positioning approaches estimate TOA from sampled signals, and the accuracy of TOA estimation is consequently influenced by the sampling rate of mobile devices.
Since the precision of TDOA-based positioning ultimately depends on the accuracy of TOA estimation, PRS-based positioning methods are crucial to minimize TOA estimation errors \cite{ref46y,ref47y}. 
However, some of the recent mobile devices, including low-complexity IoT devices, face challenges in achieving sufficient TOA estimation accuracy due to their low sampling rate. \cite{ref48y,ref49y}
In contrast to typical 5G mobile devices with high sampling rates, low-complexity mobile devices cannot acquire high-rate samples of PRS signals, limiting their ability to precise position \cite{ref50y,ref51y}.

In this paper, we apply a frequency-domain phase-based TOA estimation method, inspired by the orthogonal phase of arrival (OPA) principle, to the 5G PRS structure. The proposed algorithm estimates OPA by analyzing phase variations across orthogonal subcarriers in order to capture fine-grained residual timing offsets. This approach leverages the linear phase progression induced by fractional timing misalignments in OFDM systems, allowing sub-sample resolution without increasing the sampling rate.
While such phase-based estimation techniques have been widely studied in OFDM synchronization, we demonstrate their applicability to 5G PRS and evaluate their performance using an SDR-based testbed. In particular, we investigate how different PRS resource allocation patterns influence TOA estimation accuracy. The key contributions of this paper are summarized as follows:
\begin{enumerate}
    \item We apply the novel concept of using OPA for TOA estimation. This concept enables us to obtain Residual TOA (RTOA), corresponding to the TOA within a sample. This distinguishes the proposed algorithm from conventional studies, whose estimation accuracy is directly related to the sampling rate.
    \item We investigate the impact of PRS configurations on positioning accuracy in 5G NR. Based on the review of the configuration parameters, we analyze how the adjustment of each parameter affects the accuracy of TOA estimation. The analysis provides insights into the reception characteristics of various 5G PRS configurations. This also helps to select the best configuration parameters suited to a specific signal environment.
    \item We verify the performance of the proposed TOA estimation algorithm in real-world signal propagation environments. The proposed algorithm is implemented on a software modem that operates with USRPs that transmit and receive PRS signals. This verification provides practical performance metrics and serves as a reference for the setup of the system and experimental procedures.
\end{enumerate}

The remainder of this paper is organized as follows. 
Section \uppercase\expandafter{\romannumeral2} describes the basic mathematical model for TOA estimation, describing assumptions about the PRS signal and the structure of the frame. 
Section \uppercase\expandafter{\romannumeral3} presents a TOA estimation algorithm for higher accuracy.
Section \uppercase\expandafter{\romannumeral4} explains the testbed and experimental setup for evaluating the TOA estimation algorithm with real signal emission and provides an analysis of the experimental results. 
Finally, Section \uppercase\expandafter{\romannumeral5} concludes the paper.


\begin{table}[t]
\centering
\caption{Description of abbreviations}
\label{tab:my-table}
\begin{tabular}{|l|p{5.5cm}|}
\hline
\textbf{Abbreviation} & \textbf{Description} \\ \hline
OFDM       & Orthogonal Frequency-Division Multiplexing \\ \hline
FFT / IFFT & Fast Fourier Transform / Inverse FFT       \\ \hline
PRS        & Positioning Reference Signal               \\ \hline
TOA        & Time Of Arrival                            \\ \hline
ITOA       & Integer Time Of Arrival                    \\ \hline
RTOA       & Residual Time Of Arrival                   \\ \hline
ICI        & Inter-Carrier Interference                 \\ \hline
USRP       & Universal Software Radio Peripheral        \\ \hline
LOS        & Line Of Sight                              \\ \hline
PSD        & Power Spectral Density                     \\ \hline
QPSK       & Quadrature Phase Shift Keying              \\ \hline
MSE        & Mean Squared Error                         \\ \hline
CFR        & Channel Frequency Response                 \\ \hline
SNR        & Signal-to-Noise Ratio                      \\ \hline
\end{tabular}
\end{table}

\begin{table}[t]
\centering
\caption{Time-Frequency PRS Mapping Pattern Representing \( k' \)}
\label{tab:table2}
\resizebox{\linewidth}{!}{%
\begin{tabular}{|c|c|c|c|c|c|c|c|c|c|c|c|c|}
\hline
$K_{\text{comb}}^{\text{PRS}}$ &
\multicolumn{12}{c|}{Symbol number within the DL PRS resource $\ell - \ell_{\text{start}}^{\text{PRS}}$ (\( k' \))} \\ \hline
 &
  0 & 1 & 2 & 3 & 4 & 5 & 6 & 7 & 8 & 9 & 10 & 11 \\ \hline
2 &
  0 & 1 & 0 & 1 & 0 & 1 & 0 & 1 & 0 & 1 & 0 & 1 \\ \hline
4 &
  0 & 2 & 1 & 3 & 0 & 2 & 1 & 3 & 0 & 2 & 1 & 3 \\ \hline
6 &
  0 & 3 & 1 & 4 & 2 & 5 & 0 & 3 & 1 & 4 & 2 & 5 \\ \hline
12 &
  0 & 6 & 3 & 9 & 1 & 7 & 4 & 10 & 2 & 8 & 5 & 11 \\ \hline
\end{tabular}%
}
\end{table}

\section{System Model} \label{systemModel}
We assume a 5G PRS system model based on OFDM, consisting of a single PRS transmitter and receiver. The system employs a frame structure in which each frame comprises multiple slots, and each slot consists of multiple OFDM symbols. Each OFDM symbol contains \( N_{\text{FFT}} \) subcarriers, and \( N_{\text{SC}} \) subcarriers are grouped into a Resource Block (RB). Additionally, the system is assumed to allocate up to a maximum of \( N_{\text{RB}} \) RBs. The subcarrier spacing, denoted by \( f_{\text{SCS}} \), defines the frequency interval between adjacent subcarriers.

The PRS transmitter generates the \( m \)-th PRS sequence, denoted as \( r(m) \), as follows~\cite{ref52y}:
\begin{equation}  
    r(m) = \frac{1}{\sqrt{2}} \left[ (1 - 2c(2m)) + j (1 - 2c(2m+1)) \right],
\end{equation}
,where \( N_{\text{PRS}} \) represents the length of the PRS sequence, \( m = 0, 1, \dots, N_{\text{PRS}} - 1 \), and \( c(n) \) is the 31st-order Gold pseudo-random sequence.

The sequence \( r(m) \) is mapped onto the radio resource at the \( k \)-th subcarrier in the \( l \)-th OFDM symbol, denoted by \( X^{(l)}[k] \), as follows:
\begin{equation}  
X^{(l)}[k_m] = \beta_{\text{PRS}} r(m), \label{X[k_m]}
\end{equation}  
where \( \beta_{\text{PRS}} \) and \( k_m \) represent the power factor and the subcarrier index for the \( m \)-th PRS symbol, respectively.

The factor \( \beta_{\text{PRS}} \) determines the overall signal energy contributed by the PRS sequence, denoted as \( E_{\text{PRS}} \), as follows:
\begin{equation}
E^{\text{PRS}} = N_{\text{PRS}} \beta_{\text{PRS}}^2 = \frac{N_{\text{SC}} N_{\text{RB}}}{K_{\text{comb}}^{\text{PRS}}} \beta_{\text{PRS}}^2. \label{E_PRS}
\end{equation}

The subcarrier index \( k_m \) is determined based on configuration parameters \( K_{\text{comb}}^{\text{PRS}} \in \{ 2, 4, 6, 12 \} \) and \( k_{\text{offset}}^{\text{PRS}} \in \{ 0, 1, \dots, K_{\text{comb}}^{\text{PRS}} - 1 \} \), which denote the subcarrier spacing and the starting subcarrier index for PRS, respectively, as follows:
\begin{equation}  
k_m = m \cdot K_{\text{comb}}^{\text{PRS}} + \left( (k_{\text{offset}}^{\text{PRS}} + k') \bmod K_{\text{comb}}^{\text{PRS}} \right),
\end{equation}
where \( k' \) is determined based on Table~\ref{tab:table2}.

After resource allocation, the transmitter generates the \( l \)-th OFDM symbol, denoted by \( x^{(l)}[n] \), using OFDM modulation. Let \( L_{\text{PRS}} \in \{ 2, 4, 6, 12 \} \) denote the number of OFDM symbols containing PRS, and let \( l_{\text{start}}^{\text{PRS}} \) denote the index of the starting OFDM symbol.

The symbol \( x^{(l)}[n] \) is generated by performing an \( N_{\text{FFT}} \)-point IFFT on \( X^{(l)} = [X^{(l)}[0], \cdots, X^{(l)}[N_{\text{FFT}}-1]]^T \) for \( l = l_{\text{start}}^{\text{PRS}}, l = l_{\text{start}}^{\text{PRS}} + 1, \dots, l_{\text{start}}^{\text{PRS}} + L_{\text{PRS}} - 1 \). For subcarriers not assigned to PRS, \( X^{(l)}[k] \) is set to zero or handled according to system configuration. The overall transmitted signal, denoted by \( x[n] \), can then be expressed as:
\begin{equation}
\begin{aligned}
x[n] &= \sum_{l=0}^{L_{\text{PRS}} - 1} x^{(l + l_{\text{start}}^{\text{PRS}})}[n - l(N_{\text{FFT}} + N_{\text{CP}})], \\
&\quad \text{where } x[n] = 0 \text{ for } n > N_{\text{FFT}} - 1.
\end{aligned}
\label{eq:xn}
\end{equation}
where \( N_{\text{CP}} \) represents the length of the Cyclic Prefix (CP).

The received signal, denoted by \( y[n] \), is a continuous time-domain signal that encompasses all OFDM symbols. It can be expressed as:
\begin{equation}
y[n] = h[n] * x[n] + w[n], \label{y[n]}
\end{equation}
where \( h[n] \) and \( w[n] \) represent the Channel Impulse Response (CIR) and additive noise, respectively. Assuming a Line-of-Sight (LOS) delayed channel, \( h[n] \) can be written as:
\begin{equation}
h[n] = \alpha \delta[n - \tau], \label{LOS}
\end{equation}
where \( \alpha \) denotes the channel gain and \( \tau \) is the time delay.

The \( l \)-th received OFDM symbol, denoted by \( y^{(l)}[n] \), is extracted from \( y[n] \) based on a timing offset \( \hat{t} \), which is estimated using a synchronization algorithm. The receiver performs an auto-correlation between \( y[n] \) and \( x[n] \) as follows:
\begin{equation}
\hat{t} = \underset{t \in \Delta}{\arg\max} \, R[t], \label{corr1}
\end{equation}  
\begin{equation}
R[t] = \left| \sum_{n=0}^{N-1} y[n - t] \cdot x^*[n] \right|, \label{corr2}
\end{equation}  
where \( N \) denote the lengths of \( x[n] \) and \( y[n] \). The timing offset is estimated by computing the correlation at intervals of \( m \), where \( m \) is a positive integer. The set \( \Delta \) consists of discrete time shifts \( 0, m, 2m, 3m, \dots \), up to the maximum permissible shift within the received signal length.

We denote \( \hat{t} \) in~(\ref{corr1}) as the Integer Time of Arrival (ITOA), which is obtained in units of samples. Based on the ITOA, the \( l \)-th received OFDM symbol \( y^{(l)}[n] \) is extracted as follows:
\begin{equation}
y^{(l)}[n] = y[n + l(N_{\text{FFT}}+C_{CP}) + \hat{t}], \quad n = 0, 1, \dots, N_{\text{FFT}} - 1. \label{sync}
\end{equation}

Let \( Y^{(l)}[k] \) denote the FFT of \( y^{(l)}[n] \). Then,
\begin{equation}
Y^{(l)}[k] = H[k] X^{(l)}[k] + W^{(l)}[k], \label{CFR}
\end{equation}
where \( H[k] \) and \( W^{(l)}[k] \) represent the Channel Frequency Response (CFR) and additive noise at the \( k \)-th subcarrier of the \( l \)-th OFDM symbol, respectively.

Based on~(\ref{LOS}), the expression in~(\ref{CFR}) can be further refined as:
\begin{equation}
Y^{(l)}[k] = \alpha e^{-j 2\pi k \tau / N} X^{(l)}[k] + W^{(l)}[k], \label{Y[k]}
\end{equation}
which highlights the phase rotation induced by the time delay at the \( k \)-th subcarrier. Although \( Y^{(l)}[k] \) is already compensated for the integer timing offset \( \hat{t} \), the residual timing gap caused by low sampling resolution remains as a fractional offset, which appears as the phase rotation term associated with \( \tau \) in this expression.

\section{Residual TOA Estimation Based on Orthogonal Phase Arrival}
\vspace{0.2cm}
Positioning with ITOA faces an inherent limitation due to arbitrary and fine-grained timing offsets smaller than the sampling period. This limitation leads to estimation errors of up to one sampling period, significantly degrading positioning accuracy. 

To overcome this challenge, we propose a novel TOA estimation method that incorporates an algorithm for estimating the RTOA, defined as the difference between the ITOA and the actual TOA. The proposed method leverages OPA, which is influenced by the timing offset and captures the differences in phase shifts occurring across individual subcarriers. OPA contributes to enhancing TOA estimation precision even under sampling rate constraints. 

It enables estimation of the fractional part of the TOA and has a significant impact on positioning accuracy when combined with conventional ITOA-based positioning.

\begin{figure}
    \centering
    \includegraphics[width=1\linewidth]{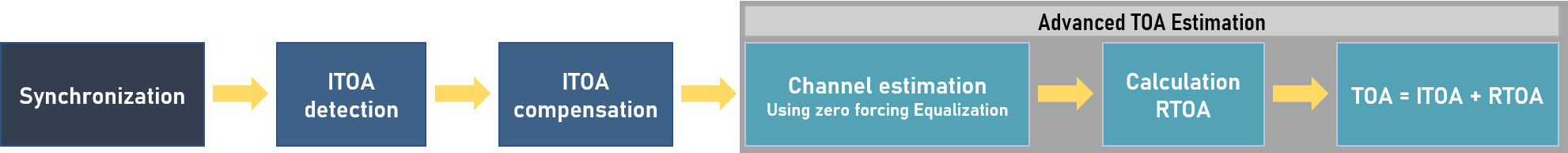}
    \caption{Flow chart of the proposed TOA estimation method.}
    \label{flowchart}
\end{figure}

The algorithmic flow of the proposed method is illustrated in Fig.~\ref{flowchart}. The process begins with time and frequency synchronization based on the synchronization signal in the received signal. Once synchronization is achieved, the ITOA is acquired through the correlation described in (\ref{corr1}) and (\ref{corr2}). Based on the obtained timing offset, the received signal is transformed into frequency-domain symbols according to (\ref{sync}) and (\ref{CFR}). The channel response is then estimated from these frequency-domain symbols. Using the estimated channel response, the proposed method derives the RTOA for refining the overall TOA calculation. By utilizing the estimated RTOA, the proposed method determines the TOA with accuracy within one sample.

\subsection{Estimation of Residual Time of Arrival}
We define \( \epsilon = t - \hat{t} \) as the Residual Time of Arrival (RTOA), representing the difference between the actual TOA \( t \) and the estimated integer TOA \( \hat{t} \). Note that \( \tau \) in~(\ref{Y[k]}) was used to describe the general time delay for explaining phase rotation, and is conceptually aligned with \( \epsilon \) in this context. Based on the resource mapping assumed in Section~\ref{systemModel}, PRS is allocated to every \( K_{\text{comb}}^{\text{PRS}} \)-th subcarrier. The proposed algorithm leverages the OPA concept by deriving the representative phase difference based on the phase shift observed between two PRS symbols with interval \( M \). Specifically, the algorithm selects two PRS symbols at the \( k \)-th and the \( (k + M K_{\text{comb}}^{\text{PRS}}) \)-th subcarriers from one OFDM symbol, and calculates the phase difference of the corresponding PRS symbols.

The estimation of the phase differential with respect to \( M \), denoted by \( \Delta \theta^{(M)} \), is mathematically expressed as follows:
\begin{align}
\Delta \theta^{(M)} &= -2\pi f_{\text{SCS}} (k + M K_{\text{comb}}^{\text{PRS}}) \epsilon + 2\pi f_{\text{SCS}} k \epsilon \nonumber \\
                    &= -2\pi f_{\text{SCS}} M K_{\text{comb}}^{\text{PRS}} \epsilon. \label{thetaM}
\end{align}

Based on (\ref{thetaM}), \( \epsilon \) can be derived from \( \Delta \theta^{(M)} \) as:
\begin{equation}
\epsilon = -\frac{\Delta \theta^{(M)}}{2\pi f_{\text{SCS}} M K_{\text{comb}}^{\text{PRS}}}. \label{RTOA1}
\end{equation}

RTOA in sample units, denoted by \( \epsilon_s \), is converted from \( \epsilon \) as:
\begin{align}
\epsilon_{s} &= -\frac{\Delta \theta^{(M)}}{2\pi f_{\text{SCS}} M K_{\text{comb}}^{\text{PRS}}} \cdot \frac{1}{T_{s}} \quad \text{[samples]} \nonumber \\
&= -\frac{N_{\text{FFT}}}{2\pi K_{\text{comb}}^{\text{PRS}}} S \quad \text{[samples]} \label{RTOA_s}
\end{align}
,where \( S = \frac{\Delta \theta^{(M)}}{M} \). The value \( S \) is the mathematical definition of OPA and represents the phase shift amount derived from the phase variation between arbitrarily selected subcarrier indices.

Deriving \( \Delta \theta^{(M)} \) requires knowledge of \( H[k] \), which is estimated using a zero-forcing approach. The estimated CFR, denoted by \( \hat{H}[k] \), is obtained as:
\begin{equation}
\hat{H}[k] = \frac{Y[k]}{X[k]}. \label{channelEstimator}
\end{equation}

To derive \( S \) with low computational complexity, the proposed algorithm divides the subcarriers into two sets, 
as at least two points are required to calculate the slope. The sets of subcarrier indices, denoted by \( K_{\text{LOW}} \) and \( K_{\text{HIGH}} \), are defined as:
\begin{align}
K_{\text{LOW}} &= \left\{ k_0 + a K_{\text{comb}}^{\text{PRS}} \mid 0 \leq a < \frac{N_{\text{PRS}}}{2} \right\}, \\
K_{\text{HIGH}} &= \left\{ k_0 + a K_{\text{comb}}^{\text{PRS}} \mid \frac{N_{\text{PRS}}}{2} \leq a < N_{\text{PRS}} \right\},
\end{align}
,where \( a \) is an integer and \( k_0 \) is the index of the first PRS subcarrier.

Using the subcarriers in each set, the proposed algorithm derives representative phase values. The averaged phase values for \( K_{\text{LOW}} \) and \( K_{\text{HIGH}} \), denoted by \( \hat{H}_{\text{LOW}} \) and \( \hat{H}_{\text{HIGH}} \), are calculated as:
\begin{align}
\hat{H}_{\text{LOW}} &= \left| \hat{H}_{\text{LOW}} \right| e^{j \theta_i^{(M)}} = \frac{1}{|K_{\text{LOW}}|} \sum_{k \in K_{\text{LOW}}} \hat{H}[k], \label{Hlow} \\
\hat{H}_{\text{HIGH}} &= \left| \hat{H}_{\text{HIGH}} \right| e^{j \theta_f^{(M)}} = \frac{1}{|K_{\text{HIGH}}|} \sum_{k \in K_{\text{HIGH}}} \hat{H}[k], \label{Hhigh}
\end{align}
,where \( |\cdot| \) denotes the cardinality of the set, and \( \theta_i^{(M)} \) and \( \theta_f^{(M)} \) denote the phases of \( \hat{H}_{\text{LOW}} \) and \( \hat{H}_{\text{HIGH}} \), respectively. This averaging process facilitates robust RTOA estimation by reducing the noise present in the received PRS symbols and further enhances the accuracy of OPA estimation by capturing and mitigating fine-scale phase distortions across subcarriers.

Using \( \hat{H}_{\text{LOW}} \) and \( \hat{H}_{\text{HIGH}} \), \( S \) is estimated as:
\begin{equation}
S = \frac{\angle (\hat{H}_{\text{HIGH}} \hat{H}_{\text{LOW}}^*)}{M}. \label{slope}
\end{equation}
\( \angle(\cdot) \) denotes the angle operator.

The proposed algorithm finally derives RTOA by substituting into (\ref{RTOA_s}) as:
\begin{equation}
\epsilon_s = - \frac{N_{\text{FFT}}} {2\pi M K_{\text{comb}}^{\text{PRS}}} \angle (\hat{H}_{\text{HIGH}} \hat{H}_{\text{LOW}}^*) \quad \text{[sample]}. \label{epsilon}
\end{equation}

The proposed algorithm estimates OPA by analyzing phase variations across orthogonal subcarriers to capture fine-grained timing offsets. It derives the phase difference \( S \) across orthogonal PRS subcarriers based on the phase shift \( \Delta \theta^{(M)} \) and enables more accurate TOA estimation. Furthermore, the proposed algorithm primarily consists of linear operations, allowing for low computational complexity.

\subsection{Algorithmic Flow Design}
\begin{figure}
    \centering
    \includegraphics[width=8.5cm]{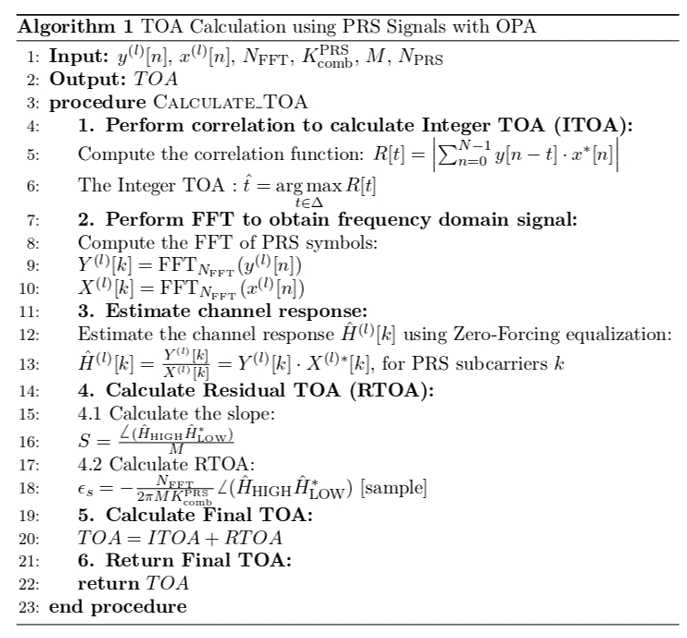}
    \caption{Pseudo-code of the proposed RTOA estimation method.}
    \label{pseudo}
\end{figure}

Based on the proposed algorithm, we design the overall TOA estimation procedure as summarized in Fig.~\ref{pseudo}. The proposed method estimates TOA by sequentially calculating the integer and fractional parts, incorporating OPA in the fractional estimation stage to enhance TOA accuracy. This ensures that OPA plays a central role in estimating phase shifts across orthogonal subcarriers, effectively mitigating timing errors.

ITOA is initially derived from correlation with the received signal, as described in (\ref{corr1})–(\ref{corr2}). Then, frequency-domain symbols are obtained by applying an FFT to the received PRS signal, aligned based on the estimated ITOA. From these frequency-domain symbols, the CFR of each PRS subcarrier is estimated using (\ref{channelEstimator}). The proposed method subsequently estimates the phase difference caused by the RTOA by calculating \( S \), as defined in (\ref{slope}), leveraging OPA for more accurate phase alignment and timing refinement. This is performed by computing \( H_{\text{LOW}} \) and \( H_{\text{HIGH}} \) using the estimated CFR, based on (\ref{Hlow}) and (\ref{Hhigh}). Finally, the method calculates the RTOA using (\ref{RTOA_s}) and estimates the overall TOA as the sum of ITOA and RTOA, explicitly integrating OPA throughout the estimation process.

The proposed algorithm effectively mitigates TOA errors in the time domain by applying OPA in the frequency domain. This contributes to achieving positioning accuracy within a single sample. Moreover, the proposed method enables precise positioning on low-complexity terminals with low sampling rates, benefiting from the inherent advantages of OPA applied to orthogonal subcarrier phase processing.

On the other hand, a limitation of the proposed method is its sensitivity to noise due to the nature of OPA. Since the RTOA estimation relies on the received signal, the accuracy of the estimation is influenced by the noise level. As the noise increases, the estimated channel phase becomes more distorted, potentially degrading the overall TOA estimation accuracy.

\section{Testbed Experiments for Validating Estimation Accuracy}
\begin{figure}
    \centering
    \includegraphics[width=0.7\linewidth]{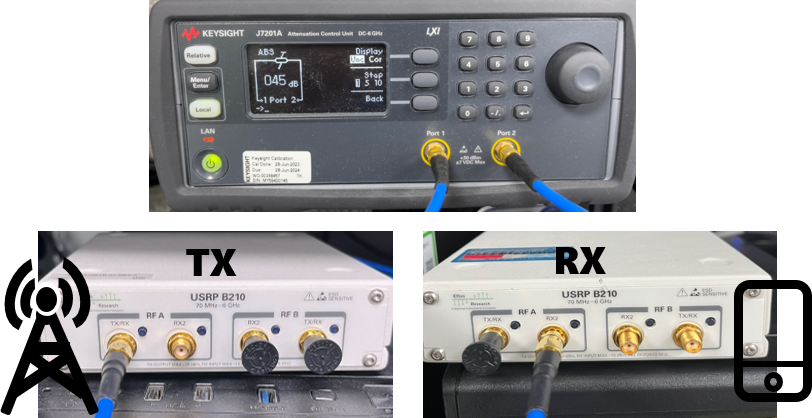}
    \caption{USRP antenna connection ports for PRS transmission and reception.}
    \label{RXTX}
\end{figure}

To evaluate the contribution of the proposed OPA technique to estimation accuracy, the algorithm is implemented on a testbed system. 
As illustrated in Fig.~\ref{RXTX}, the testbed consists of a 5G PRS transmitter and receiver employing USRP B210 devices for signal transmission and reception. 
The received PRS signal power is controlled via a signal attenuator placed between the transmitting and receiving USRPs. 
Baseband signal processing is performed by a software modem operating on Linux-based PCs.
To emulate an AWGN environment with a LOS-like channel, the transmitter and receiver are connected through an RF conduction cable and signal attenuator.

\begin{figure}
    \centering
    \includegraphics[width=0.8\linewidth]{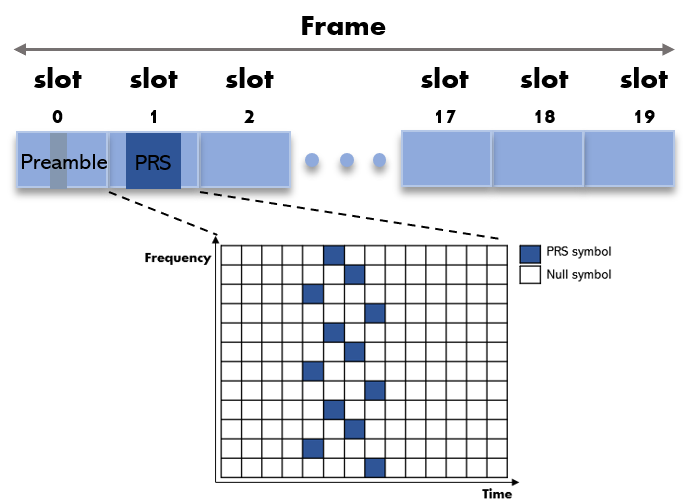}
    \caption{5G PRS frame structure used in the experiment.}
    \label{framestructure}
\end{figure}

\begin{table}[h!]
\centering
\footnotesize
\caption{Simulation parameters.}
\label{parameter}
\resizebox{0.7\columnwidth}{!}{
\begin{tabular}{|c|c|}
\hline
\textbf{Parameter}                 & \textbf{Value}                     \\ \hline
FFT size                           & 1024                               \\ \hline
Subcarrier spacing                 & 30~kHz                              \\ \hline
Sampling rate                      & 30.72~MHz                           \\ \hline
Center frequency                   & 3.3~GHz                             \\ \hline
Preamble sequence                  & M-sequence                          \\ \hline
PRS sequence                       & Gold sequence                       \\ \hline
\end{tabular}
}
\end{table}

The experiment assumes a frame structure compliant with commercial 5G systems, as depicted in Fig.~\ref{framestructure}. 
Each 10~ms frame consists of 20 slots in the time domain, with each slot containing 14 OFDM symbols. 
The preamble and PRS are transmitted in slot 0 and slot 1, respectively. 
Before processing the PRS signal, the receiver performs coarse time and frequency offset compensation using the preamble. 

PRS resource allocation is configured with the parameters \( L_{\text{PRS}} = 4 \), \( l_{\text{start}}^{\text{PRS}} = 4 \), \( K_{\text{comb}}^{\text{PRS}} = 4 \), \( k_{\text{offset}}^{\text{PRS}} = 1 \), and \( N_{\text{RB}} = 20 \), aiming to minimize synchronization errors that could distort OPA estimation and degrade TOA accuracy.

Table~\ref{parameter} summarizes the detailed testbed configuration. 
The subcarrier spacing is set to 30~kHz, consistent with commercial 5G deployments. 
The sampling rate is 30.72~MHz, in alignment with LTE configurations. 
A center frequency of 3.3~GHz is chosen to avoid interference with commercial FR1 band signals. 
Both the preamble and PRS sequences follow 5G NR standard configurations.

Experiments are conducted under varying PRS power allocation settings based on the defined configuration. 
The number of PRS symbols per OFDM symbol is adjusted using \( K_{\text{comb}}^{\text{PRS}} \) and \( N_{\text{RB}} \), following (\ref{E_PRS}), while the energy per PRS symbol is controlled via \( \beta_{\text{PRS}} \) in (\ref{X[k_m]}).

The accuracy of TOA estimation is assessed by computing the mean squared error (MSE) of the estimated TOAs, obtained by averaging the receiver’s estimates and comparing them against ground-truth TOAs.

\subsection{Functional Assessment of OPA-Based TOA Estimation}


\begin{figure}
    \centering
    \includegraphics[width=0.8\linewidth]{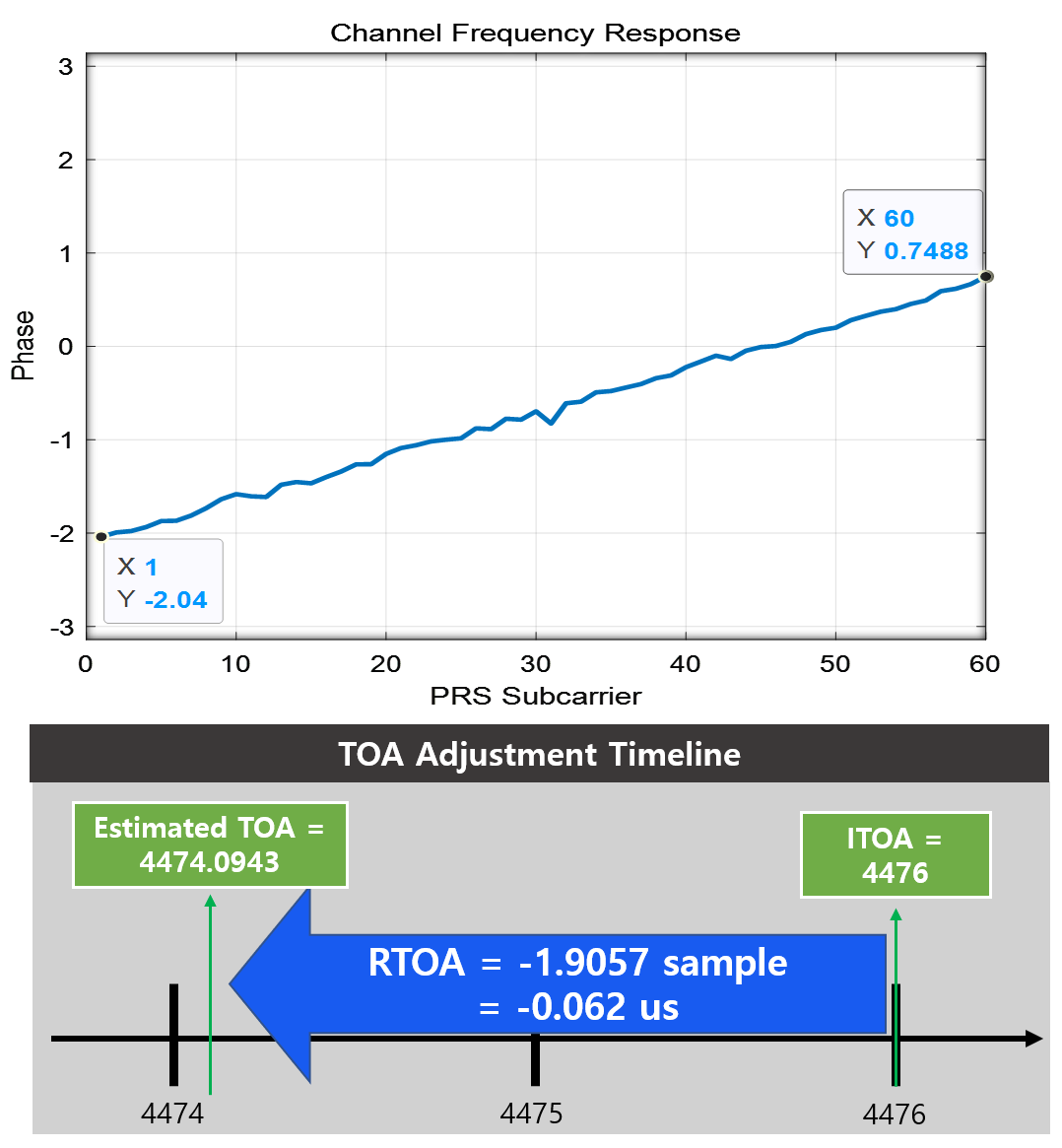}
    \caption{Channel Frequency Response (Top) and TOA Adjustment Timeline (Bottom)}
    \label{CFR&rtoa}
\end{figure}
\begin{figure}
    \centering
    \includegraphics[width=1\linewidth]{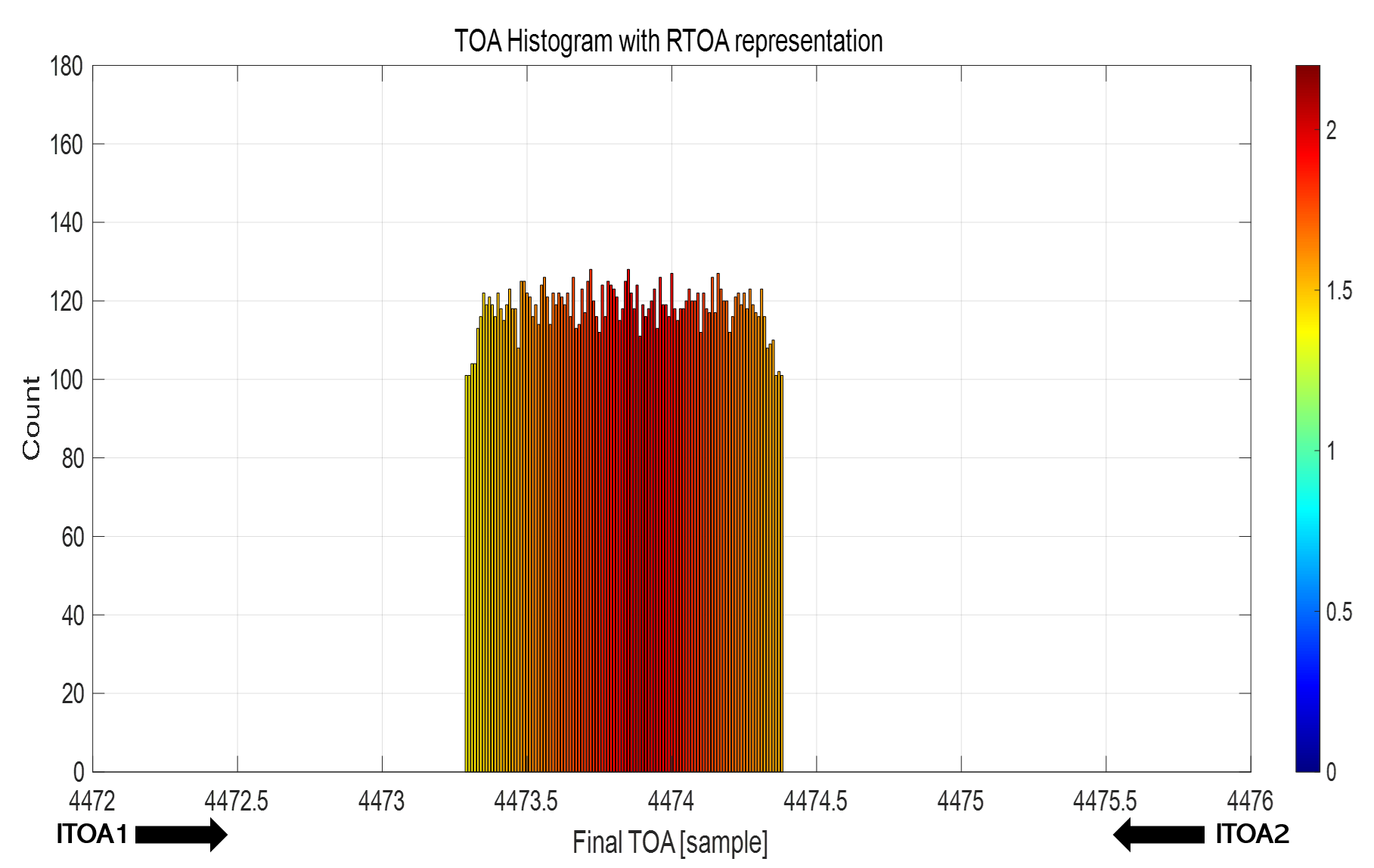}
    \caption{Final TOA distribution with RTOA representation}
    \label{distribution}
\end{figure}

 We first observe the instantaneous operation of the proposed algorithm from a functional perspective. 
We capture the samples of the received signal using the USRP and apply the proposed algorithm to estimate the TOA in MATLAB. 
The ITOA is estimated as 4476 samples, which corresponds to the configured timing of the PRS in terms of \( l_{\text{start}}^{\text{PRS}} \). 
The proposed method acquires the OPA based on the ITOA to estimate the timing difference between the actual TOA and the ITOA. 
As illustrated in Fig.~\ref{CFR&rtoa}, the phase of the CFR exhibits a linear trend, where the slope \( S \) is observed as 0.0467 under the assumption that \( M = 60 \). 
Based on (\ref{epsilon}), the proposed method derives the RTOA as \(-1.905\)\footnote{A negative value indicates that the ITOA exceeds the actual TOA.} samples, equivalent to \(-0.062~\mu\text{s}\), and obtains the final estimated TOA as 4474.0995 samples by compensating for the ITOA.

Fig.~\ref{distribution} statistically demonstrates how the proposed method provides RTOA through the histogram of estimated TOA values. 
The estimated TOAs exhibit a uniform distribution in the range of 4473.5 to 4474.5, as TOA is successively estimated by tracking sample-level timing offsets using preamble signals. 
This uniformity results from timing drift caused by minor sampling clock mismatches and frequency offsets that occur during the test. 
The mean of the estimated TOAs is 4474, which corresponds to the target timing offset aimed by the timing compensation algorithm. 
Therefore, this statistical observation supports that the RTOA estimation performed by the proposed method is both valid and practically reliable. 
Furthermore, the results confirm that the TOA can be estimated with a resolution finer than one sample, which ultimately contributes to enhanced positioning accuracy. Since the experiment is not simulation-based, the actual ground truth of TOA is unknown, and thus absolute accuracy cannot be directly assessed. Nonetheless, the tight distribution around the mean provides insight into the precision of the proposed method.

 \subsection{Performance of TOA Estimation Based on Orthogonal Phase Arrival}

\begin{figure}
    \centering
    \includegraphics[width=1\linewidth]{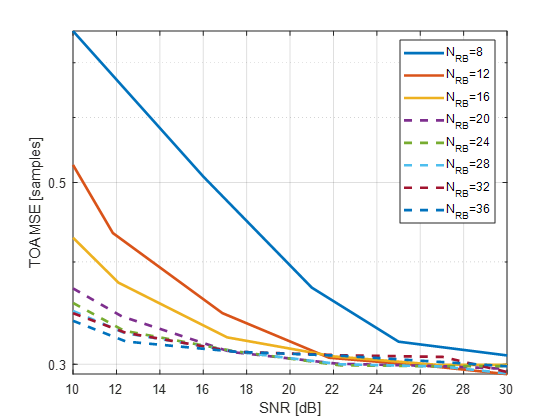}
    \caption{The MSE of TOA estimation according to resource block}
    \label{plot1}
\end{figure}

Fig.~\ref{plot1} illustrates the MSE of the estimated TOA under various signal environments and resource block allocations to PRS. 
The results show that the TOA estimation error generally converges below a certain threshold as the SNR increases. 
When \(N_{\text{RB}} \leq 12\), the MSE remains greater than 1 sample even at an SNR of 20~dB, indicating limited estimation accuracy with such a resource allocation. 
In contrast, allocating 24 or more \(N_{\text{RB}}\) significantly improves estimation performance, reducing the MSE to below 0.3 samples. 
Specifically, at an SNR of 20~dB, the MSE decreases from approximately 1.4 samples to about 0.29 samples (equivalent to 13.5~m and 2.8~m) as \(N_{\text{RB}}\) increases from 8 to 24. 
A marginal performance improvement is also observed beyond 24 RBs, as increasing \(N_{\text{RB}}\) from 24 to 32 results in an MSE reduction of less than 0.02 samples.

 The analysis of the results presented in Fig.~\ref{plot1} provides insights into the minimal resource allocation required to ensure stable TOA estimation. 
In this configuration, allocating 24 RBs to PRS guarantees reliable TOA estimation while avoiding excessive resource usage. 
When \( N_{\text{RB}} \geq 24 \), the MSE converges to approximately 0.3 samples, equivalent to around 2.8 meters, which largely satisfies the accuracy requirements for indoor positioning as defined by 3GPP~\cite{ref53y}. 
Therefore, an allocation of at least 24 RBs can be considered a practical threshold that balances estimation accuracy with resource efficiency in TOA-based positioning systems.

\begin{figure}
    \centering
    \includegraphics[width=1\linewidth]{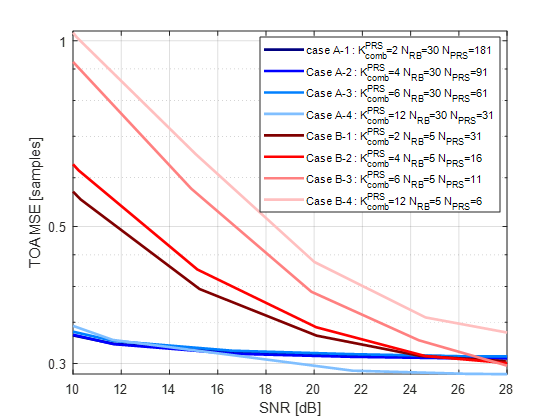}
    \caption{The MSE of TOA estimation according to PRS subcarrier count}
    \label{plot2}
\end{figure}
The effect of PRS allocation patterns, in terms of subcarrier interval and the number of resource blocks, on TOA estimation accuracy is shown in Fig.~\ref{plot2}. 
The results aim to compare two representative scenarios: one with wideband resource allocation and the other with narrowband allocation. 
The wideband allocation, denoted as Case A, assumes that sufficient resources are allocated to PRS, whereas the narrowband allocation, denoted as Case B, represents a bandwidth-constrained environment.
Such an environment is exemplified by use cases like 5G Reduced Capability (RedCap) UEs, which operate with limited bandwidth and are targeted for low-cost, low-power IoT applications.
Both cases are evaluated across various subcarrier intervals, with all other parameters held constant.



For Case A, increasing \(K_{\text{comb}}^{\text{PRS}}\) has minimal impact on estimation accuracy. 
This is because the number of PRS subcarriers remains sufficient to suppress noise, even at the maximum \(K_{\text{comb}}^{\text{PRS}}\) value. 
In contrast, Case B exhibits strong sensitivity of estimation accuracy to \(K_{\text{comb}}^{\text{PRS}}\). 
In a lower SNR regime, a smaller \(K_{\text{comb}}^{\text{PRS}}\) yields a significant reduction in MSE, as the PRS allocation becomes denser and allows for greater averaging in (\ref{Hlow}) and (\ref{Hhigh}).

This comparison underscores the importance of subcarrier interval selection under bandwidth-limited conditions and provides broader insights into PRS design strategies under such constraints. 
With sufficient bandwidth, \(K_{\text{comb}}^{\text{PRS}}\) can be selected flexibly, as even a large subcarrier interval ensures an adequate number of PRS subcarriers for reliable estimation. 
However, when the bandwidth allocated to PRS is limited, increasing the subcarrier interval significantly reduces the number of usable PRS subcarriers, resulting in notable degradation in estimation performance. 
To this end, the selection of \(K_{\text{comb}}^{\text{PRS}}\) should be carefully adapted to the available frequency resources to maintain robust TOA estimation.




\begin{figure}
    \centering
    \includegraphics[width=1\linewidth]{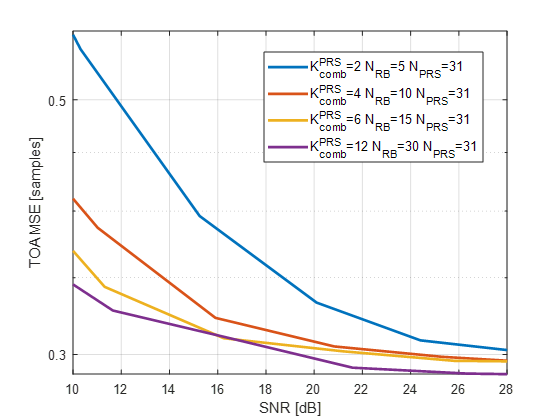}
    \caption{The MSE of TOA estimation according to PRS subcarrier interval}
    \label{plot3}
\end{figure}

Fig.~\ref{plot3} presents the MSE performance of TOA estimation for various PRS allocation patterns using a fixed number of PRS subcarriers. 
Compared to Case B in Fig.~\ref{plot2}, all four cases in Fig.~\ref{plot3} show a consistent decrease in MSE as \(K_{\text{comb}}^{\text{PRS}}\) increases. 
This indicates that smaller PRS subcarrier intervals increase inter-carrier interference (ICI) between adjacent PRS subcarriers, thereby degrading TOA estimation accuracy. 
Due to residual timing and frequency offsets, a certain level of ICI is inevitable and becomes more pronounced between closely spaced PRS subcarriers, especially as \(K_{\text{comb}}^{\text{PRS}}\) increases, leading to distortion in the received symbols~\cite{ref54y}.

\begin{figure}
    \centering
    \includegraphics[width=1\linewidth]{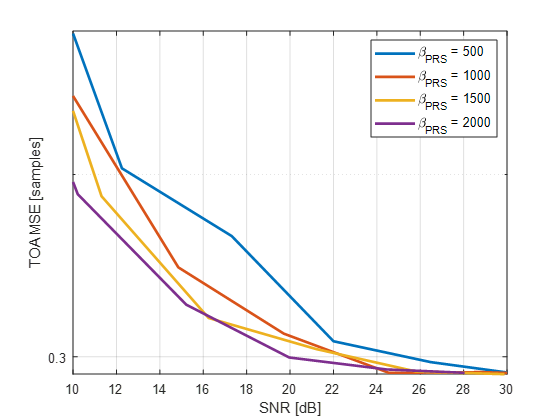}
    \caption{The MSE of TOA estimation according to energy per PRS subcarrier}
    \label{plot4}
\end{figure}

 
Fig.~\ref{plot2} and Fig.~\ref{plot3} highlight how PRS allocation parameters should be determined in accordance with radio resource availability. 
When traffic is low and radio resources are abundant, wideband PRS allocation—with both a large subcarrier interval and a high number of resource blocks—is desirable. 
Under this configuration, TOA estimation remains robust even with a large \(K_{\text{comb}}^{\text{PRS}}\), owing to the sufficient number of PRS subcarriers and reduced ICI. 
In contrast, under high traffic conditions with limited radio resources, PRS must be allocated to a smaller number of resource blocks. 
In such cases, a large \(K_{\text{comb}}^{\text{PRS}}\) leads to performance degradation due to the reduced number of PRS subcarriers, while a small \(K_{\text{comb}}^{\text{PRS}}\) provides better estimation accuracy despite increased ICI. 
These observations suggest that PRS allocation should be adaptively configured according to traffic and bandwidth conditions. 
Using wide PRS subcarrier intervals is effective for minimizing ICI when resources are sufficient, whereas employing denser PRS patterns helps preserve signal quality under resource-constrained environments.


 Fig.~\ref{plot4} presents the TOA estimation performance in terms of the transmit power of a PRS subcarrier. 
The results show that increasing \(\beta_{\text{PRS}}\) enhances the received energy of PRS symbols and consistently reduces the MSE under a given SNR condition. 
However, beyond a certain point, further increase in \(\beta_{\text{PRS}}\) yields only marginal improvements in MSE. 
This indicates that the TOA estimation accuracy reaches saturation when the received signal power exceeds a certain threshold, implying that simply increasing PRS symbol power does not always lead to improved TOA estimation accuracy.


\begin{figure}[t]
    \centering
    \includegraphics[width=0.8\linewidth]{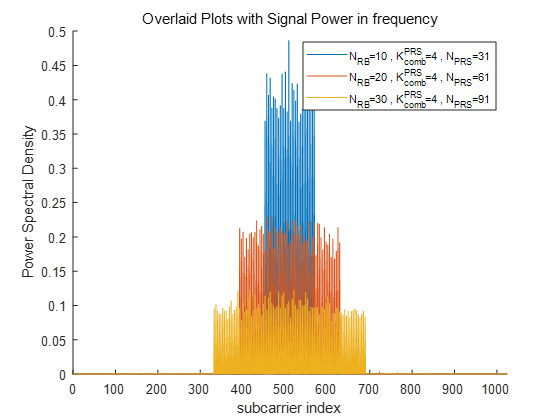}
    \caption{Power spectral density of PRS signals corresponding to the configurations in Fig. 12}
    \label{plot5}
\end{figure}

\begin{figure}[t]
    \centering
    \includegraphics[width=1\linewidth]{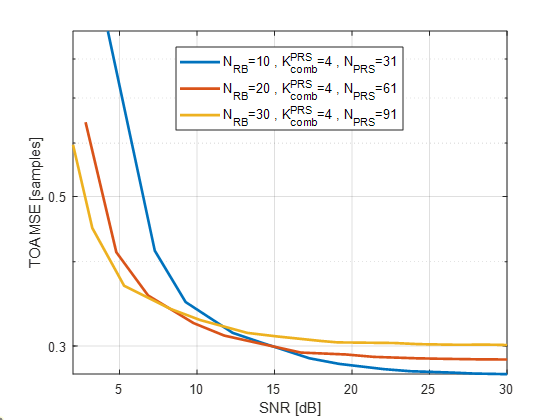}
    \caption{The MSE of TOA estimation according to PRS pattern length (via resource block variation)}
    \label{plot6}
\end{figure}


We also conduct experiments with various PRS allocation patterns under a fixed \(E^{\text{PRS}}\). 
The three PRS allocation patterns illustrated in Fig.~\ref{plot5} exhibit that the transmit power of a PRS subcarrier is inversely proportional to \(N_{\text{PRS}}\). 

Fig.~\ref{plot6} presents the TOA estimation performance for each PRS allocation pattern. 
The allocation with 10 RBs yields the lowest MSE in the high SNR regime, as the signal strength of each received PRS subcarrier is sufficiently high to support stable TOA estimation. 
Conversely, the allocation with 30 RBs demonstrates more robust performance in the low SNR regime. 
This result indicates that increasing the PRS symbol density across the frequency domain enhances the averaging gain and helps to mitigate the impact of noise.

 
The observations from Fig.~\ref{plot3} through Fig.~\ref{plot6} suggest that PRS resources should be adaptively allocated according to the target signal environments. 
When the positioning service targets users with relatively high SNR, such as those in dense urban scenarios, it is efficient to allocate fewer RBs to PRS and assign higher energy to each subcarrier. 
This strategy not only achieves high estimation accuracy but also reduces both resource usage and the computational load of the estimation process. 
Conversely, when the positioning service targets users in more challenging environments with lower SNR, it is more beneficial to allocate PRS across a larger number of RBs to ensure robust estimation, even at the cost of reduced energy per subcarrier.

\begin{figure}[!t]
    \centering
    \includegraphics[width=0.8\linewidth]{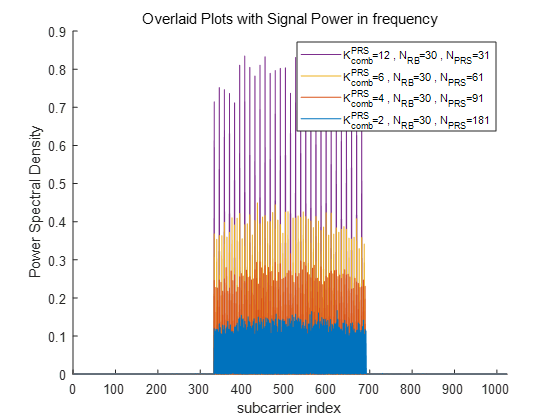}
    \caption{Power spectral density of PRS signals corresponding to the configurations in Fig. 14}
    \label{plot7}

    \vspace{0.5cm} 

    \includegraphics[width=1\linewidth]{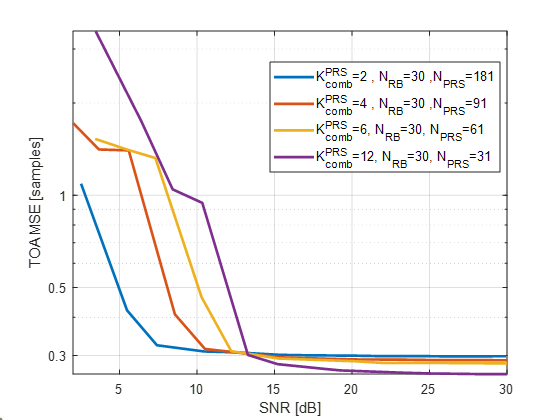}
    \caption{The MSE of TOA estimation according to PRS density (subcarrier allocation)}
    \label{plot8}
\end{figure}


To analyze the impact of the PRS subcarrier interval on TOA estimation performance, we conducted experiments by varying \(K_{\text{comb}}^{\text{PRS}}\) while keeping \(N_{\text{RB}}\) and \(E^{\text{PRS}}\) constant. 
As \(K_{\text{comb}}^{\text{PRS}}\) decreases, \(N_{\text{PRS}}\) increases, and the energy per PRS subcarrier decreases accordingly, as illustrated in Fig.~\ref{plot7}. 
As shown in Fig.~\ref{plot8}, a larger subcarrier interval results in a lower MSE when the SNR exceeds 15~dB, thereby ensuring more stable TOA estimation. 
This indicates that strong received power from PRS subcarriers effectively mitigates TOA estimation errors, even with limited averaging gain due to the reduced number of subcarriers. 
In contrast, when the SNR is below 15~dB, a larger subcarrier interval leads to a significant increase in MSE. 
Meanwhile, PRS allocations with smaller subcarrier intervals exhibit a more gradual increase in MSE, attributed to the greater averaging gain provided by a larger number of subcarriers.

The results suggest that the PRS subcarrier interval should also be adaptively configured according to the signal environment. 
In high-SNR environments, configuring a large \(K_{\text{comb}}^{\text{PRS}}\) enables higher symbol energy and is advantageous for improving TOA estimation accuracy. 
In contrast, in low-SNR environments, a small \(K_{\text{comb}}^{\text{PRS}}\) provides better noise suppression and contributes to more stable estimation. 
This highlights a fundamental trade-off between the number of PRS subcarriers and the energy per subcarrier, which should be carefully balanced during resource allocation.

\section{Conclusion}

In this study, we proposed a novel TOA estimation method using OPA to enhance positioning accuracy in low sampling rate environments. The proposed method achieves sub-sample resolution by leveraging the RTOA estimated from the phase of the channel frequency response. It was evaluated in various 5G PRS-based environments, and the experimental results confirmed that TOA estimation performance is significantly influenced by PRS resource allocation patterns. In particular, the number of PRS subcarriers, subcarrier spacing, and energy per subcarrier jointly affect estimation accuracy under different SNR conditions.

These findings demonstrate the practical feasibility of achieving high-precision positioning while utilizing radio resources efficiently. The results presented in Section~V provide practical guidelines for optimizing PRS resource allocation based on traffic and signal conditions. As future work, the proposed method can be extended to utilize other types of reference signals, such as sounding reference signals (SRS) or demodulation reference signals (DMRS), and can be combined with beamforming strategies to further enhance estimation accuracy. Additionally, adaptive resource allocation under dynamic channel conditions or in multi-cell scenarios can be explored to improve the robustness and applicability of the proposed method.

\section*{Conflict of interest}
The authors declare that there is no conflict of interest in this paper.

\bibliographystyle{IEEEtran}
\bibliography{references}

\begin{IEEEbiography}[{\includegraphics[width=1in,height=1.25in,clip,keepaspectratio]{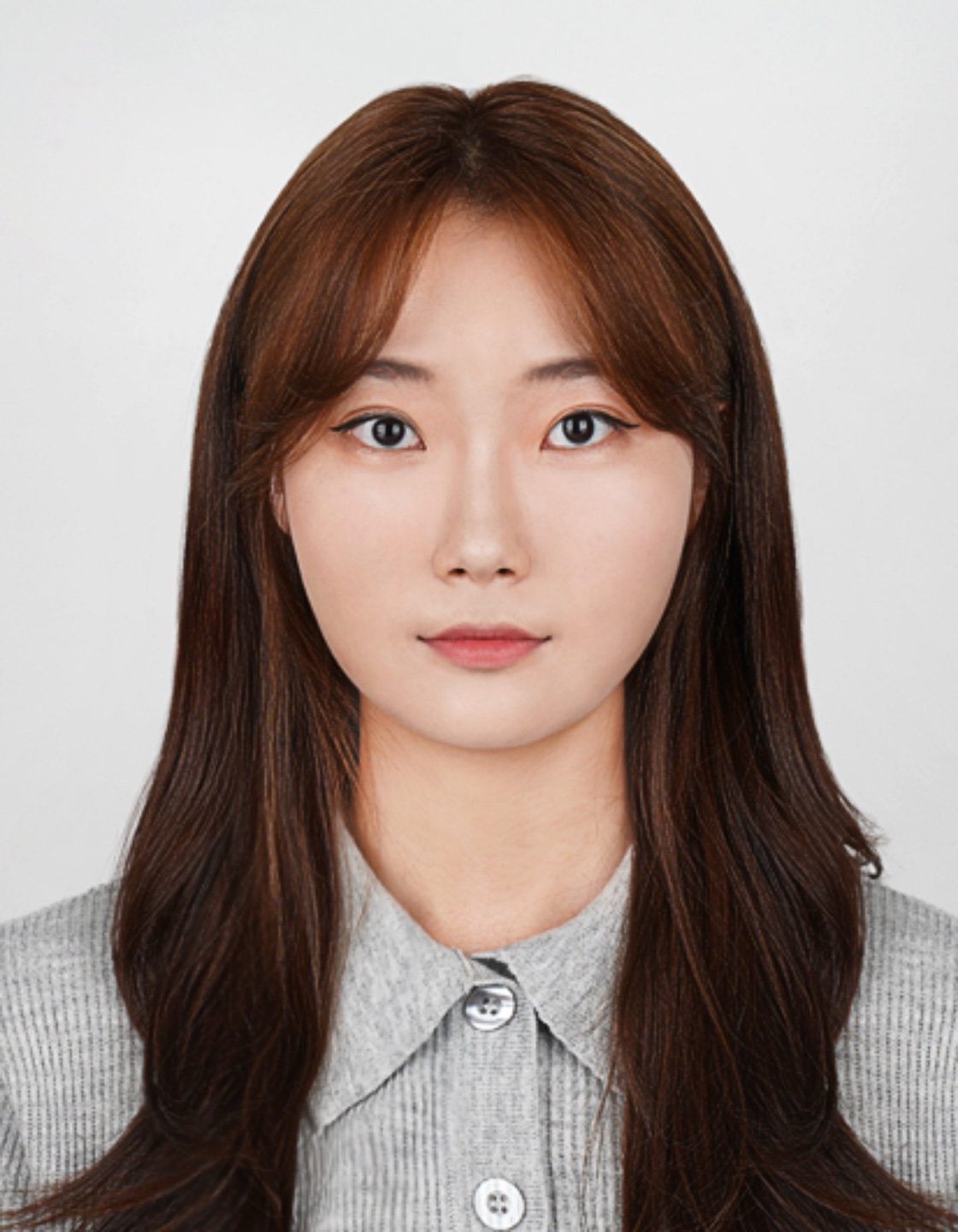}}]{Hyejin Shin}
received the B.S. degree in electronics engineering from Sookmyung Women’s University, Seoul, Korea, in 2023. She is currently pursuing the M.S. degree in Electronics Engineering at Sookmyung Women’s University. Her research interests include signal processing and software modem development.
\end{IEEEbiography}

\begin{IEEEbiography}[{\includegraphics[width=1in,height=1.25in,clip,keepaspectratio]{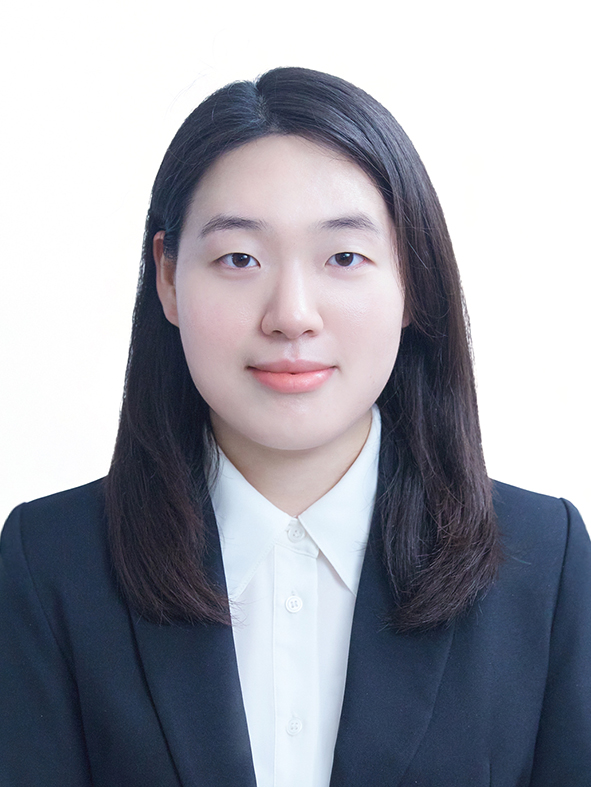}}]{Sohee Kim}
received the B.S. degree in electronics engineering from Sookmyung Women’s University, Seoul, Korea, in 2023. She is currently pursuing the M.S. degree in Electronics Engineering at Sookmyung Women’s University. Her research interests include signal processing and software modem development.
\end{IEEEbiography}

\begin{IEEEbiography}{Juyeop Kim}
(S'06--M'10) is an assistant professor in the Department of Electronics Engineering, Sookmyung Women's University, Seoul, Korea. He received the B.S. and Ph.D. degrees in electrical engineering from KAIST in 2004 and 2010, respectively. His research interests include software modems and next-generation wireless communication.
\end{IEEEbiography}

\begin{IEEEbiography}{Ilmu Byun}
is with the Korea Railroad Research Institute, Uiwang, South Korea. His research interests include wireless communications and railway applications.
\end{IEEEbiography}

\EOD
\end{document}